\documentclass[aps,twocolumn,groupedaddress]{revtex4}
\usepackage{graphicx}
\begin{document}
\title{Identifying ``communities'' within energy landscapes} 
\author{Claire P. Massen}
\affiliation{University Chemical Laboratory, Lensfield Road, Cambridge CB2 1EW, United Kingdom}
\author{Jonathan P.~K.~Doye}
\email{jpkd1@cam.ac.uk}
\affiliation{University Chemical Laboratory, Lensfield Road, Cambridge CB2 1EW, United Kingdom}
\date{\today}

\begin{abstract}

Potential energy landscapes can be represented as a network of minima linked by transition states.
The community structure of such networks has been obtained for a series of small Lennard-Jones clusters.
This community structure is compared to the concept of funnels in the potential energy landscape.
Two existing algorithms have been used to find community structure, one involving removing edges with high betweenness, the other involving optimization of the modularity.
The definition of the modularity has been refined, making it more appropriate for networks such as these where multiple edges and self-connections are not included.
The optimization algorithm has also been improved, using Monte Carlo methods with simulated annealing and basin hopping, both often used successfully in other optimization problems.
In addition to the small clusters, two examples with known heterogeneous landscapes, LJ$_{13}$ with one labelled atom and LJ$_{38}$, were studied with this approach.
The network methods found communities that are comparable to those expected from landscape analyses.
This is particularly interesting since the network model does not take any barrier heights or energies of minima into account.
For comparison, the network associated with a two-dimensional hexagonal lattice is also studied and is found to have high modularity, thus raising some questions about the interpretation of the community structure associated with such partitions.

\end{abstract}

\maketitle

\section{Introduction}

Networks have been the focus of much attention in the last few years \cite{Albert02,Newman03,Dorogovtsev03}.
This is partly due to the growth of two networks---the internet \cite{Faloutsos99} and the world wide web \cite{Albert99}.
A diverse range of networks have also been studied, for example interaction networks \cite{Jeong01,Wagner01} of proteins or genes, social networks such as the actor network \cite{Barabasi99,Albert00b} and networks of collaborations amongst scientists \cite{Newman01c}.
Initially, the focus was on the common features that many of these networks possessed, such as their small-world character \cite{Watts98} or whether the degree distribution was scale-free \cite{Barabasi99}.

The current emphasis is on more detailed properties of such networks, in particular to understand how these properties differ between different types of networks and how these differences reflect the origins of the networks.
Properties that have been studied include correlations between connected nodes \cite{Newman02}, community structure \cite{Girvan02} and possible spatial embedding \cite{Gastner04,Rozenfeld02}.
The focus here is on community structure, and as highlighted in a recent summary of `the ten leading questions for network research' there are important open questions in this area \cite{tenq}.
For example, what is the best way to quantify such community structure and what are its origins?

The networks that we focus on in this work are associated with potential energy landscapes.
The potential energy of a system is a function of the coordinates of all $N_{at}$ atoms.
Plotting this would give a 3$N_{at}$-dimensional `hyper-surface' that is known as a potential energy landscape (PEL) \cite{Landscapes}.
Energy landscapes have been the focus of much attention in recent years, as scientists have sought to understand the behaviour of a system, such as the ability of a protein to fold to its native state or the super-Arrhenius slowing down of the dynamics in fragile liquids, in term of features of the landscape.

Some of the most interesting points on PELs are stationary points where the gradient vanishes.
The type of stationary points that receive the most focus are the minima since they represent locally stable structures.
It can be particularly useful to partition the PEL into basins of attraction surrounding each minimum on a PEL, where a basin of attraction is defined as the set of points for which following the steepest descent pathways from those points will lead to the same minimum.
Pioneering work by Stillinger \cite{Stillinger84} has shown that systems generally spend most of their time vibrating in a minimum's basin of attraction and occasionally hop to another basin, although at sufficiently high temperature this time scale separation begins to break down \cite{Schroder00}.
First-order saddle points (often called transition states) are particularly important to describe this hopping, since the trajectory of the system passes along a transition state valley connecting adjacent basins during this process.
Thus, the minima, their energies, their connectivities and the heights of barriers between them control the dynamics of the system, i.e.\,how it moves around the PEL, and hence how the structure changes.

The number of minima increases exponentially with the number of atoms in a system.
Therefore, for all but quite small systems, it will be impossible to completely characterize the PEL.
Instead, the aim is to obtain a statistically accurate representation of the PEL from an incomplete sample of minima and transition states.
This problem has essentially been solved for minima, allowing the distribution of minima as a function of energy to be obtained \cite{Doye95,Sciortino99,Buchner99}.
Hence, a landscape description of the thermodynamics is possible.

However, the situation is much more complex for the dynamics, because not only is the distribution of transition states required, but also their pattern of connectivities.
Particularly little is known about the latter, and the aim of our research programme is to get insights into the fundamental organization of the PEL by studying the PEL connectivity for small systems for which a complete characterization can be obtained.
To achieve this we have applied the tools developed to study complex networks to the network of minima linked by transition states.
It should be noted that this representation only describes the connectivity of the PEL, its topology.
It does not take into account the energies of the minima or the heights of the barriers between them, its topography.

Recently we showed for a series of small Lennard-Jones (LJ) clusters that the networks show both a small-world behaviour and have a power-law tail to the degree distribution \cite{Doye02,Doye04b}.
In these networks the low-energy minima with large basins of attraction act as hubs in the network.
Furthermore, a detailed characterization of a variety of topological properties, such as the local clustering coefficients, degree-degree correlations and measures of betweenness showed behaviour similar to other scale-free networks.
In this paper we extend this work by applying algorithms designed to identify communities of strongly connected nodes to our PEL networks.
Our aim is to compare the divisions of the PEL achieved by these methods, which are based on purely topological information, to methods from the field of energy landscapes that also use topographical information.

\section{Landscapes, funnels and community structure}

In social networks, people are often grouped into communities where there are particularly strong interactions between the people in a community, and the people share some common characteristics or purpose. 
Attempts to quantify these concepts in network terms usually focus on the greater density of edges within communities than between them.
For example, the algorithm of Girvan and Newman \cite{Girvan02} discussed in the following section was applied to a network where nodes were people in an organization and edges linked those that communicated by email \cite{Tyler03}; the community structure identified by the algorithm corresponded to informal social groups within the organization, as confirmed by the people involved.

In the PEL networks studied here, communities would indicate groups of minima interconnected by many transition states.
It would be reasonable to expect conversions between those minima to occur rapidly whereas those between different communities/groups of minima would occur more slowly.
Community structure would then have a strong effect on the dynamics of the system.
If there is strong community structure then groups of minima are segregated from each other, probably making it more difficult to sample all of the PEL.
However, the topography of the PEL is also expected to have a strong effect on this.
For example, a group of minima might have many transition states between them and hence be classified as a community in this approach, but if the barriers are high in energy then they are unlikely to be used by the system.

The relationship between dynamics and community structure has been put to use previously in chemistry via the master equation approach \cite{Landscapes}.
The master equation describes the rate of flow of the occupation probability between minima in terms of a matrix of transition rates between each individual minima, calculated using unimolecular rate theory.
The eigenvectors of this matrix are associated with different dynamical processes.
If the network has two different communities, the slow mode associated with transitions between those communities has an eigenvector where the sign of the components depends upon in which community the minimum is.
The main difference between that approach and this work is that the only information used here is the connectivity.
In network terms the focus is an unweighted, undirected adjacency matrix, rather than a transition matrix where the strength of the connections depends on properties such as the height of the barrier between the two minima.
Our approach is equivalent to studying the system at infinite temperature, where all barriers can be equally easily crossed.
Interestingly, spectral methods have also been used to divide networks into communities based on the eigenvalues and eigenvectors of the adjacency matrix \cite{Capocci04,Newman04}.

Simple PELs such as those of the small clusters studied here are shaped like funnels leading down to the global minimum \cite{Miller99b,Doye99b} and therefore may be expected to have no community structure.
Other systems, e.g.~the 38-atom LJ cluster, have multiple funnel structures with large barriers between the different funnels \cite{Doye99b}.
In LJ$_{38}$ the global minimum is a truncated octahedron and is at the bottom of a narrow funnel of face-centred-cubic (fcc) structures, whereas most low-energy structures are icosahedral and at the bottom of a second wider funnel.
Evidence for these two funnels comes from analyses of both the topography of the PEL and the cluster's dynamics \cite{Miller99}.
The LJ$_{38}$ disconnectivity graph, which provides a pictorial representation of the barriers between the minima, shows a clear division of the low-energy minima into two sets, termed funnels because the energy decreases as one steps down towards the low-energy minima at the funnel bottom.
There are low barriers between minima in the same funnel, but a much larger barrier between the two funnels.
Unsurprisingly given that the energy barrier is a major determinant of transition rates, there is a separation of time scales between transitions between minima within a funnel and transitions between the two funnels.
There is also a thermodynamic solid-solid transition from the fcc to icosahedral structures just below the melting point \cite{Doye99}.
There are more icosahedral structures than fcc structures, so icosahedral structures are favoured entropically at higher temperatures, whereas the low energy global minimum is favoured at low temperatures.

LJ$_{38}$ will be studied here to determine whether network methods based purely on connectivity can find groups corresponding to the two funnels.
The sample of minima and transition states we analysed was previously used to study the multiple funnel topography and the thermodynamics and dynamics of the solid-solid transitions \cite{Miller99}.
Unlike the small clusters however, this sample is far from complete.
The method previously used to identify the minima in the two funnels is based on a master equation approach.

Any structure of a cluster will have many permutational isomers, structures which are identical except that the positions of some atoms have been exchanged.
For a monoatomic cluster, the number of possible permutations of each geometrically distinct minimum is $N_{at}!$.
As some of these are related by symmetry this number is reduced to $ 2N_{at}!/o $, where $o$ is the order of the point group of that minimum.

In our analyses of the PEL connectivity of LJ clusters so far we have not made any attempt to distinguish such permutational isomers of the same minimum.
Instead, we have just looked at the connections between geometrically distinct minima.
The PELs of small LJ clusters are usually described as having a single-funnel topography.
However, if one starts to consider different permutational isomers the situation can become more complex.
For example, if for LJ$_{13}$ one considers one atom to have a different mass, there are two physically distinguishable isomers of the icosahedral global minimum, one with the heavy atom in the centre of the icosahedron and one with this atom on the surface.
As there is a significant barrier for exchange of an atom between the centre and surface of the icosahedron, the two isomers of the global minimum lie at the bottom of separate funnels \cite{Wales04}.

\section{Methods}

\subsection{Community structure algorithms}

Identifying communities within a network is a non-trivial task, because the number of possible divisions of the network is very large, especially since, in general, the sizes and number of communities are unknown.
A number of different algorithms have been proposed to attempt this task \cite{Newman04,Radicchi03,Zhou03,Capocci04,Reichardt04,Simonsen04,Wu04,Latapy04}.
A recent method proposed by Girvan and Newman (the betweenness algorithm) \cite{Girvan02,Newman03c} relies solely on the connectivity of the nodes and has proven successful in various networks.
The betweenness of an edge is based on the number of shortest paths that pass along it.
If an edge lies between two communities, it is expected to have higher betweenness because lots of shortest paths will pass through it that connect nodes in the different communities (from the definition of a community there will be few inter-community edges for the shortest paths to choose from).
If the edges between communities are removed then the network will be broken down into components corresponding to the different communities of the original network.
After each edge is removed the topology of the network changes and the betweenness must be recalculated.
The algorithm is hierarchical, edges are removed until the network is broken down from one community of $N$ nodes into $N$ communities of one node each with various community splits in between.
This can be represented in a dendrogram showing the various splits in a hierarchical manner, a vertical line at any point gives the community split at that point.

To determine which of the many splits found by this algorithm is the best, a quantitative measure of the community structure is needed.
The fraction of edges lying within communities should be large for a good community split.
This is denoted by $\sum_ce_c$ where $e_c$ is the fraction of edges lying within community $c$.
However, the maximal value of $\sum_ce_c$ is found if the whole network is considered as one community, in which case all edges lie within that community and $\sum_ce_c$=1.
To rectify this, Newman and Girvan introduced the modularity $Q$.
To calculate $Q$, the fraction of edges that would lie within the same communities in a random network where the nodes have the same degree is subtracted from $\sum_ce_c$.
Therefore, if there are more edges in communities than would be expected due to the degree distribution, $Q$ is large.
A plot of $Q$ against the number of edges removed will have a peak with typical values of $Q$ around 0.3--0.7 if there is community structure.

The fraction of edges within communities in the real network is simply counted, and the expected value for a random network is calculated from the degree of the nodes.
If an edge is picked at random, the probability $a_c$ that one end of it leads to community $c$ is proportional to the number of edges that lead there, which in turn is simply the sum of the degrees of all nodes in that community.
Therefore
\begin{equation}
a_c = \frac{\sum{_{i \in c}} k_i}{2M}.
\end{equation}
\noindent The probability that both ends of the edge lead to community $c$ is then $a_c^2$, and so $Q$ is given by
\begin{equation}
Q=\sum_c \left( e_c - a_c^2 \right).
\label{eqtn:q}
\end{equation}

Having such a measure of community structure also naturally leads to a new approach to finding communities.
Rather than just using $Q$ to evaluate the output of a community structure algorithm one can instead try to optimize $Q$ directly.
In the first application of this approach, Newman used a simple agglomerative and greedy algorithm \cite{Newman03d}.
Starting with $N$ communities each containing one node, communities are joined together one node at a time until there is a single community.
At each step, the two communities that lead to the largest increase in $Q$ (or if no increases in $Q$ are possible, the smallest decrease) are grouped together.
This greedy algorithm is faster than that based on betweenness, scaling as $\mathcal{O}((M+N)N)$ rather than $\mathcal{O}(M^2 N)$.

Finding the maximal value of $Q$ is a global optimization problem, and while the above greedy algorithm can find community splits with high modularity, there is of course no guarantee that it will find the split with maximum modularity.
Indeed, given the large search spaces for the larger networks we analyse, there is good reason to think that it will not find this optimal split.
If better community splits are sought, more sophisticated global optimization algorithms should be used.

In this paper a second approach to optimize $Q$ is also investigated.
This is based on the Monte Carlo method with simulated annealing.
At each step, a node and a community are chosen at random.
The community could be any of the existing communities, including the one that the node is already in, or it could be a new community that does not contain any nodes.
Moving the node from its initial community to the new community would change $Q$ by $\Delta Q$.
If $\Delta Q$ is greater than zero then the move is accepted, otherwise the move is accepted with probability $ \exp(\beta \Delta Q) $.
This is the Metropolis criterion, where $\beta$ represents the inverse temperature.
At high temperatures many moves are accepted and lots of different community splits are sampled, whilst at lower temperature fewer splits are sampled but those that are generally have higher modularities.
The principle of simulated annealing involves starting the algorithm at high temperature and decreasing the temperature until $Q$ becomes constant because no further moves are accepted.
The hope of such an annealing scheme is that the final partition will be the global optimum, but of course it could only be a local optimum.
To increase the likelihood of success, quenches can also be applied periodically.
In these quenches, $\Delta Q$ is calculated for moving all nodes to all communities, and the move with the highest $\Delta Q$ is taken.
This is repeated until the highest $\Delta Q$ is less than or equal to zero implying $Q$ cannot be increased further.
The simulated annealing method runs quickly, taking a similar amount of time as the greedy optimization but finding higher values of $Q$.

However, simulated annealing is often not a particularly efficient global optimization method.
Indeed, we were able to obtain still higher values of the modularity using a basin hopping approach that is known to perform well in other optimization problems \cite{Wales97}.
Basin-hopping also uses Monte Carlo, but there are a number of significant differences from the simulated annealing scheme described above.
Firstly, each step consists of randomly changing the communities of a series of nodes, not just a single one.
Secondly, after each step the new partition is quenched and the Metropolis acceptance criterion is applied to the values of the modularity for the partitions that result from quenching.
Then, if a step is accepted, the current partition is updated to that of the quenched partition.
This algorithm is slower to run to completion, but finds high values of $Q$ quickly.

The initial point for both these approaches can be any partition of the nodes into communities.
This could be $N$ communities of one node, but the algorithms are faster if the initial partition is either that obtained from the greedy algorithm or a community split obtained in a similar agglomerative manner but using Monte Carlo, which is much faster.
To create this initial partition, rather than checking all possible edges as in the greedy algorithm, a single edge is chosen at random and added if $\Delta Q$ meets the Metropolis criterion.
Simulated annealing runs were started from the partition obtained from the greedy algorithm, which gave much higher values of $Q$.
If the algorithm is started from the second more random initial partition then the results depend quite strongly on parameters such as the cooling rate and frequency of quenches.
However, for the basin hopping algorithm the initial conditions, in general, have only a weak effect on $Q_{max}$, so this was started from the more random partition.
This makes basin hopping more feasible for larger networks, because the greedy algorithm, although generally fast, especially using the algorithm of Ref. \cite{Clauset04}, requires large arrays for large networks.

\subsection{Randomized networks}

In order to interpret the results of the above algorithms, it is important to compare the values of $Q$ obtained to those for an appropriate null model.
Given the importance of the degree distribution for other network properties, the usual null model is an ensemble of random networks that have the same degree distribution as the network being analysed.
Currently, the most efficient algorithm to generate such an ensemble is the rewiring method \cite{Maslov04,Milo03}.
The starting point is the real network.
Two edges A--B and C--D are then picked at random and rewired to give either A--C and B--D or A--D and B--C.
Moves are simply not allowed if they would create multiple edges or self-connections.

\begin{figure}
\centerline{\includegraphics[width=8.6cm]{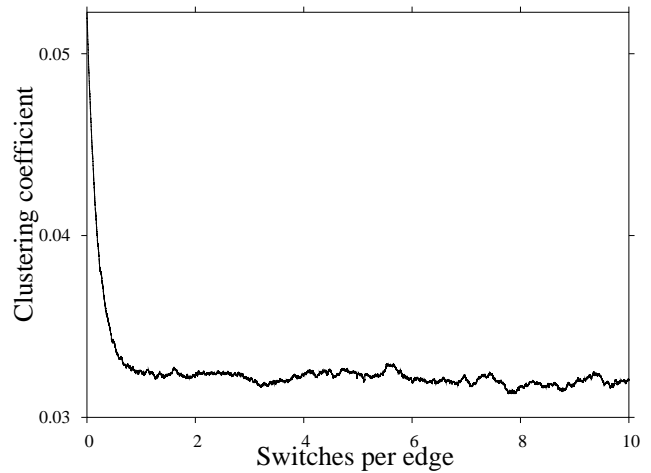}}
\caption{Variation in the clustering coefficient of the LJ$_{14}$ network as edges were rewired to form a random ensemble. Data collection began at 5 switches per edge.}
\label{fig:mc}
\end{figure}

As edges were rewired the clustering coefficient was measured to follow the randomization.
This is shown plotted against the number of rewired edges (switches) per edge of the network for LJ$_{14}$ in Figure \ref{fig:mc}.
All of these LJ networks appear to be fully randomized after around one switch per edge, so data collection for the ensemble began at five switches per edge.
It is likely that the clustering coefficient reaching its equilibrium value indicates that the networks are random after one switch per edge, but since the algorithm is fairly fast, rewiring more edges has no disadvantages.
For all random networks with between five and ten switches per edge, the probability that any two nodes were connected was recorded.
25 random networks were also saved at intervals of 0.2 switches per edge (roughly the autocorrelation time of the clustering coefficient) from the same range.
The community structure algorithms were then run on this ensemble, recording the mean and standard deviation of the modularity.

\section{Results}
\subsection{LJ$_{8-14}$}

\begin{figure}
\includegraphics[width=8.4cm]{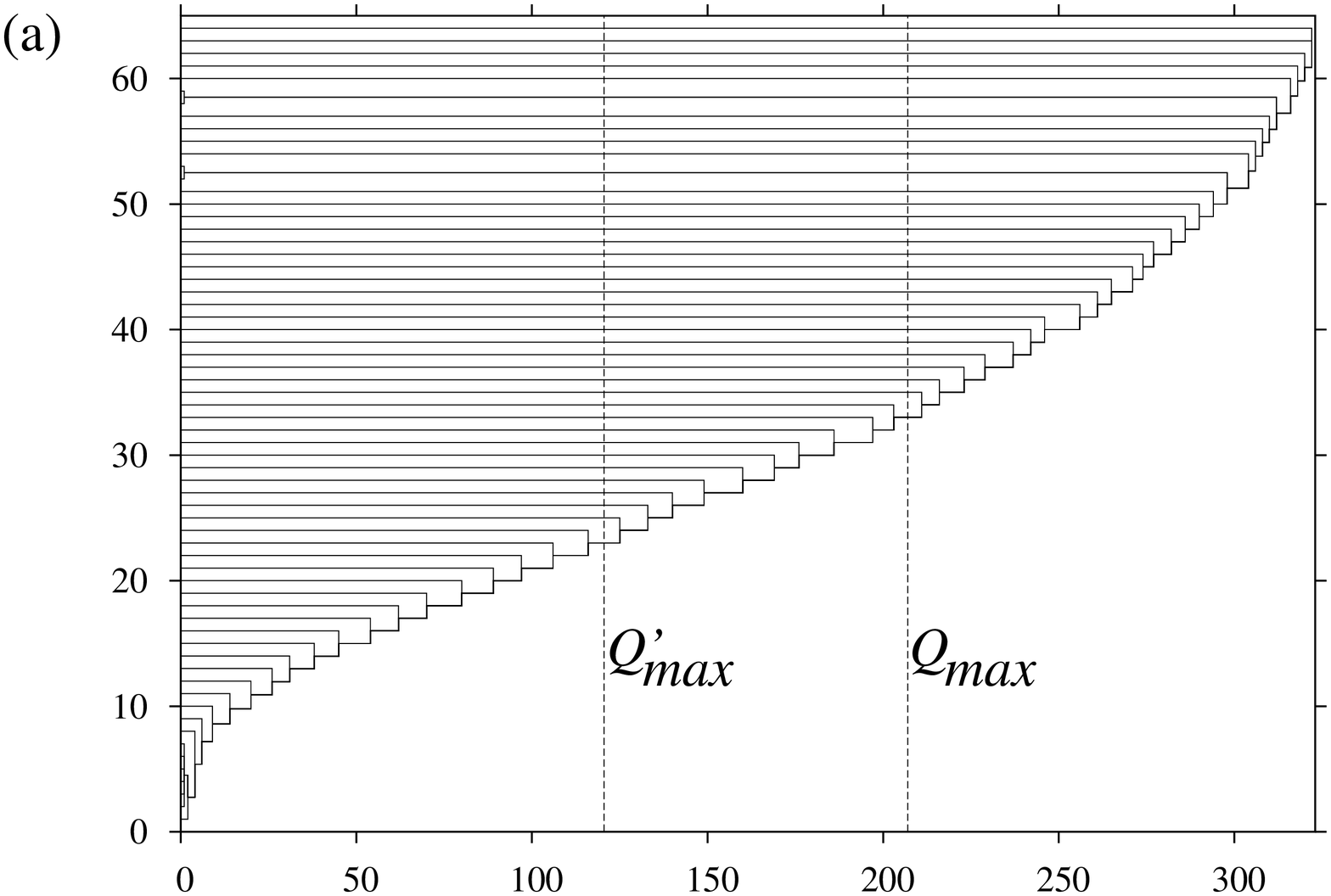}
\includegraphics[width=8.5cm]{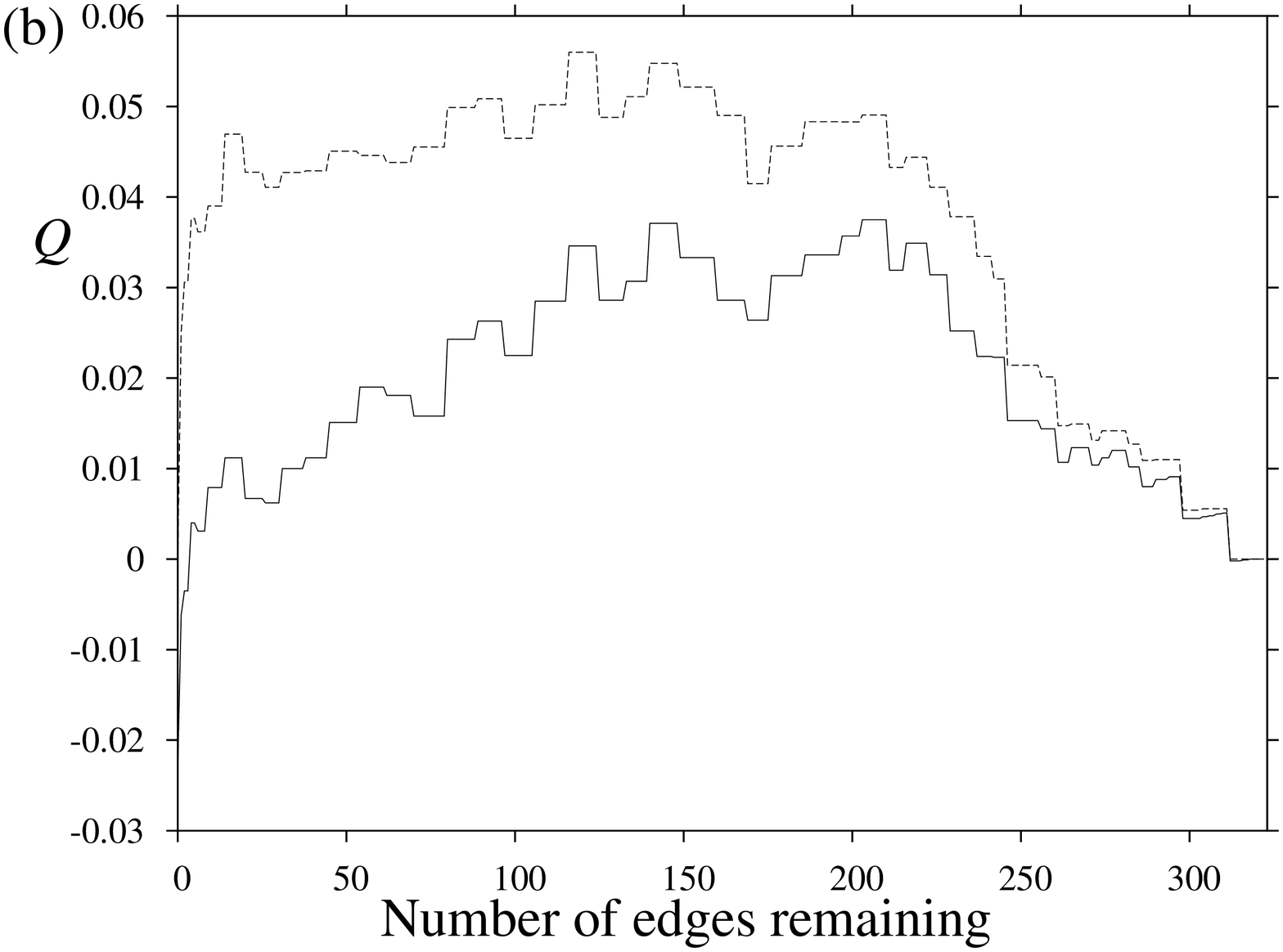}
\caption{(a) Dendrogram showing the results of the betweenness algorithm applied to LJ$_{10}$.
Each point on the y-axis represents a single node in the network (in arbitrary order).
Drawing a vertical line at any point, as has been done at the best splits as determined by $Q_{max}$ and $Q'_{max}$, gives the number and sizes of communities at that point.
In the case of the betweenness algorithm, the initial point is the right-hand side where all nodes are in a single community.
The scale of the x-axis is the number of edges remaining in the network.
(b) Variation in $Q$ (solid line) and $Q'$ (dashed line) as the algorithm progresses.}
\label{fig:bet-bothq}
\end{figure}

The results of applying the edge betweenness algorithm to identify community structure are illustrated in Figure \ref{fig:bet-bothq} by the dendrogram and the associated modularity $Q$ for LJ$_{10}$ (all small cluster networks studied showed qualitatively similar behaviour).

\begin{figure}
\centerline{\includegraphics[width=8.6cm]{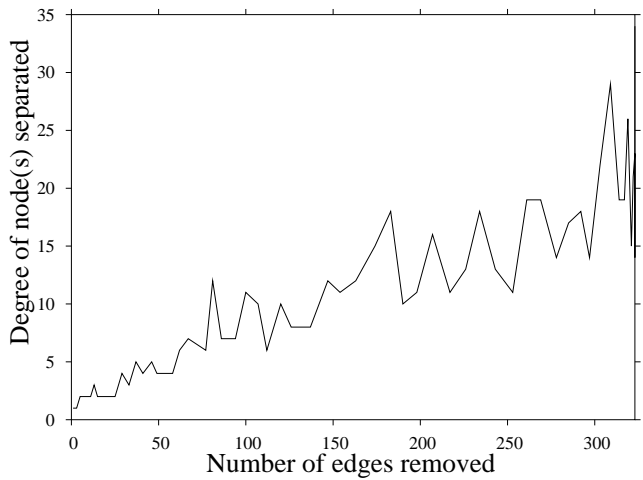}}
\caption{Degree of nodes separated from the main community by the betweenness algorithm for LJ$_{10}$.}
\label{fig:kt}
\end{figure}

The dendrogram shows that nodes were removed from the initial `community' of $N$ nodes roughly one by one.
At any stage there is one large community and several smaller `communities' consisting of one or two nodes each.
The global minimum is the last node remaining in the original community, which is perhaps unsurprising since it has the highest degree.
A plot of the degree of the node removed at each stage against the order in which they are removed (Figure \ref{fig:kt}) shows that lowest degree nodes are removed first.

This behaviour has been seen in previous work for random networks without community structure \cite{Guimera02} and our random networks also have qualitatively similar dendrograms.
In the case of a node with degree one, all shortest paths to that node must pass along the one edge connecting it to the rest of the network, making the betweenness of that edge $2(N-1)$ (the factor of 2 appears because in calculating the betweenness each path is counted twice).
This is true in any network, but the fact that these edges have the highest betweenness reflects the relatively low betweenness of the rest of the edges in the LJ networks.
This has been used previously to identify a lack of community structure in a network \cite{Wilkinson04,Tyler03}.
In that case, when enough edges had been removed to make the edges connected to degree one nodes have the highest betweenness, the algorithm was considered to have run to completion.
Since the magnitude of $Q$ is so small (an order of magnitude smaller than most networks showing community structure) the results seem to indicate that the LJ networks have little community structure.

\begin{figure}
\includegraphics[width=8.4cm]{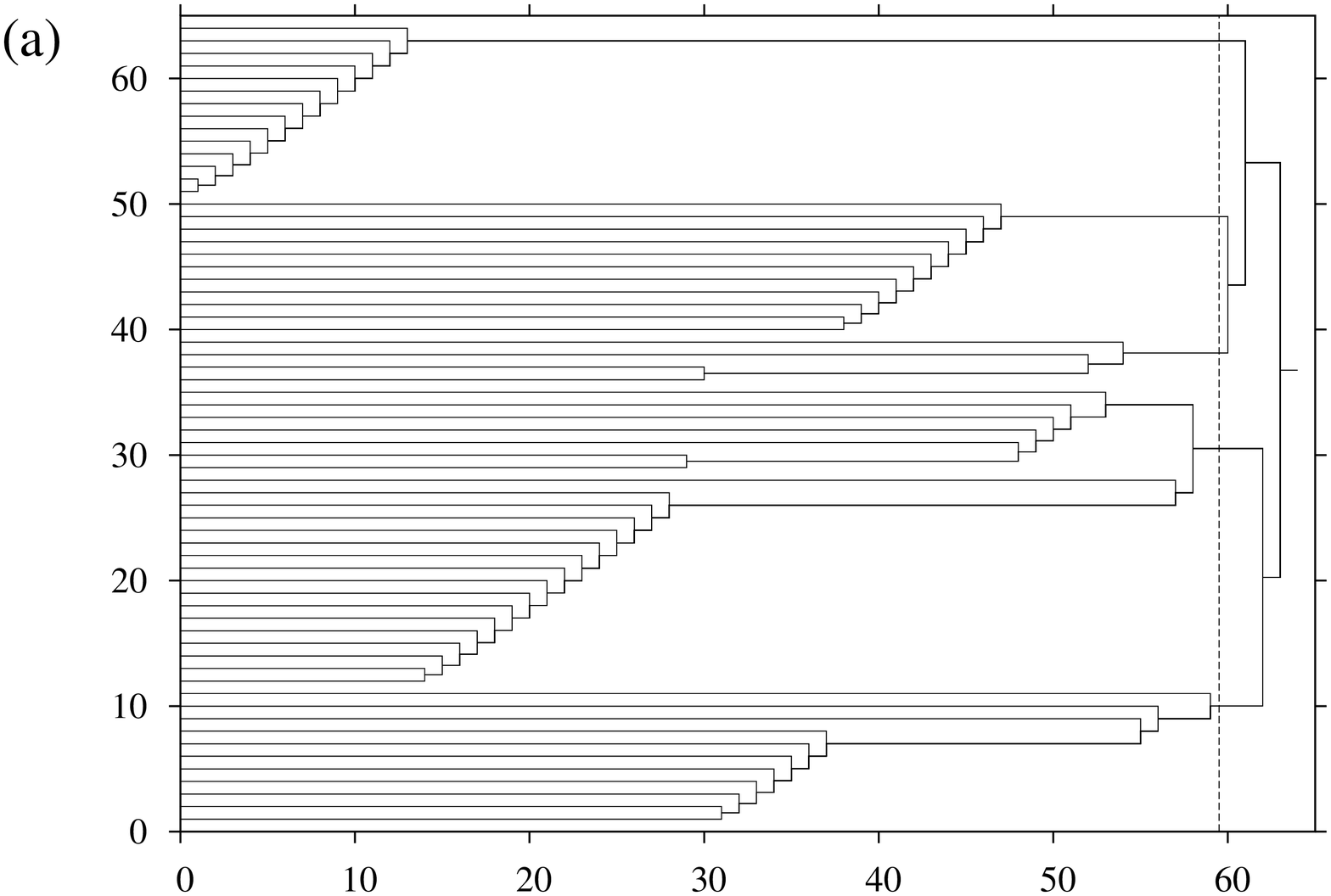}
\includegraphics[width=8.5cm]{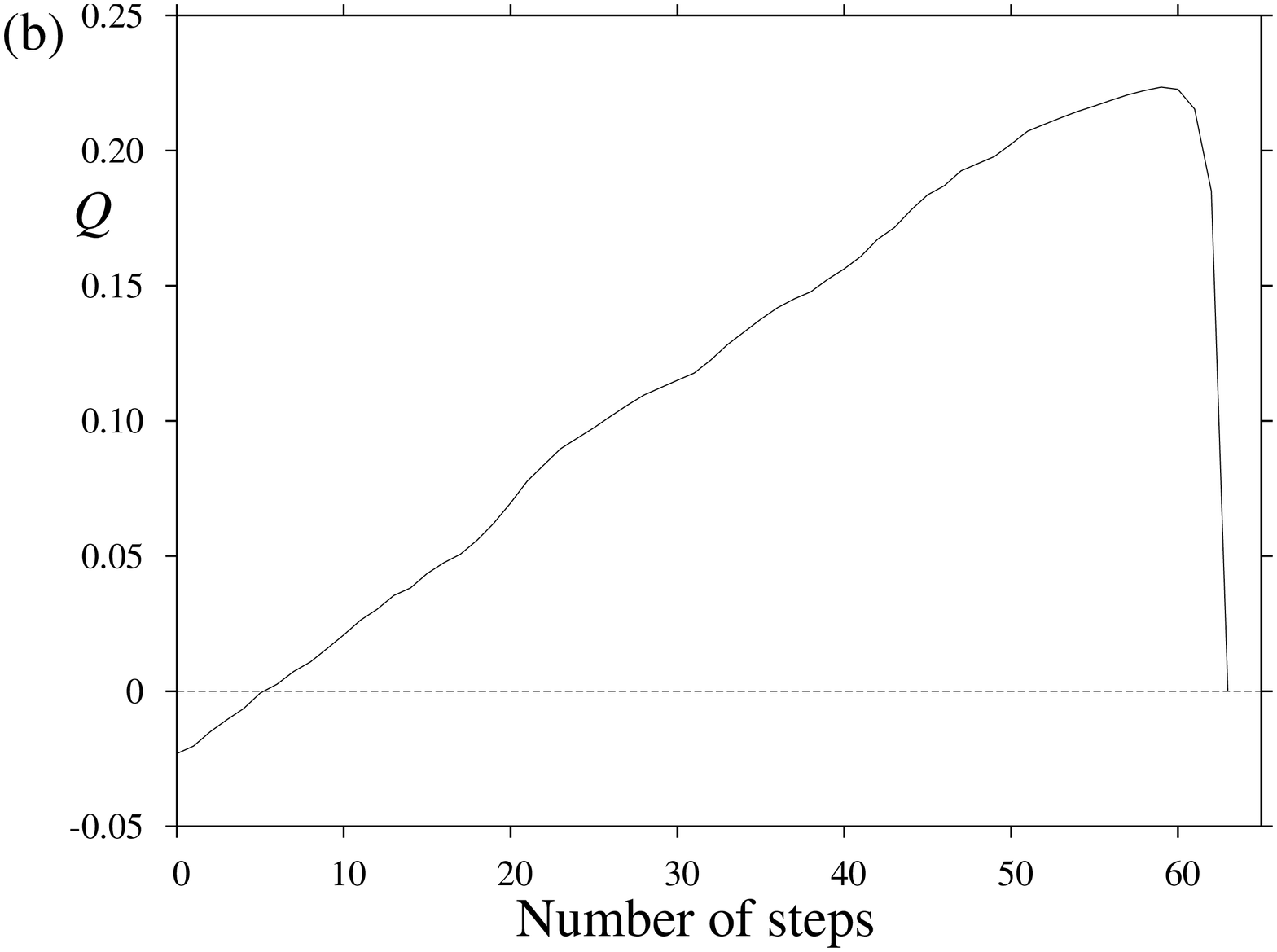}
\caption{(a) Dendrogram showing the results of the greedy algorithm applied to LJ$_{10}$.
The initial point for the greedy algorithm is at the left-hand side with $N$ (64) communities.
The x-axis represents the number of steps, where each step involves joining together two communities.
The dashed line represents the best split found (with maximum $Q$).
(b) Variation of $Q$ as the algorithm progresses.}
\label{fig:fast-oldq}
\end{figure}

Using the greedy algorithm to optimize $Q$, however, does find some community structure, as illustrated by the dendrogram for LJ$_{10}$ in Figure \ref{fig:fast-oldq}.
The community structure is weak compared to other networks---the magnitude of $Q$ for the best split, $Q_{max}$, is around 0.2--0.3.
$Q_{max}$ for the greedy algorithm is much higher than that for the betweenness algorithm, therefore the community splits from this algorithm are much better.
The betweenness algorithm has been found to be unsuccessful for dense networks (e.g.~a food web \cite{Girvan02}), since if there are many edges between communities the shortest paths between the communities can pass through any of these edges and the betweenness of any one is unlikely to be high.
It is noteworthy that in this respect the average degree of the LJ networks are significantly larger ($\langle k \rangle$ up to 30 \cite{Doye04b}) than those of many of the networks previously studied.
The betweenness algorithm has also been able to detect communities in weighted networks when it has failed for the equivalent unweighted network \cite{Newman04}.
Here we concentrate on improving the quantification of community structure and hence the community structure found by optimizing this value.

$Q$ is negative for community splits with many small communities (to the left in the dendrogram), implying that the community structure is weaker than in a random network with the same degree distribution.
The reason for this is that in `communities' of one node there can be no edges within that community.
However, the expression for $Q$ (Equation (\ref{eqtn:q})) uses a predicted fraction of edges of \( a_c^2 = \left(\sum_{i \in c} k_i/2M \right)^2 \) that is clearly greater than zero.
This effect makes $Q$ smaller when there are lots of small communities, as is the case for the betweenness algorithm, because the second term in $Q$ predicts more edges than are physically possible.
The total number of edges predicted is equal to the total number of edges in the network, so if too many are predicted in one area too few must be predicted elsewhere.
Therefore, as the communities grow this effect is reduced until $Q=0$ when there is one large community.
Using $a_c$ to predict the number of edges in a community if the network were random is equivalent to predicting $k_ik_j/2M$ edges between two nodes $i$ and $j$.
This number can clearly be greater than one for two high degree nodes.
For the largest network studied, LJ$_{14}$, the global minimum has degree 3201 and there are 61085 edges, so more than one edge is predicted between the global minimum and any node with degree greater than $\sim$\,40.
This is similar to the argument used by Maslov and Sneppen to explain degree-degree anticorrelations seen for the internet \cite{Maslov04}.
The lack of self-connections means $Q$ disfavours small communities and the lack of multiple edges means $Q$ disfavours the grouping together of high degree nodes.

A solution to this problem is to improve the predicted fraction of edges within communities.
The method used here was to create an ensemble of random networks with the same degree distribution as the original and with the constraint that multiple edges and self-connections are forbidden.
It is then straightforward to calculate the probability that there is an edge between any pair of nodes in a network taken from that ensemble.
The predicted fraction of edges within each community in a random network (with the same degree distribution and no multiple edges or self-connections) is then the sum of this probability over each pair of nodes within the community.
This sum (denoted by $f_c$) replaces $a_c^2$ in Equation (\ref{eqtn:q}) giving

\begin{equation}
Q'=\sum_{c} (e_c-f_c).
\label{eqtn:q'}
\end{equation}

\noindent It may also be possible to use the analytical method of Park and Newman \cite{Park03} to predict the probability of an edge between two nodes.

\begin{figure}
\includegraphics[width=8.6cm]{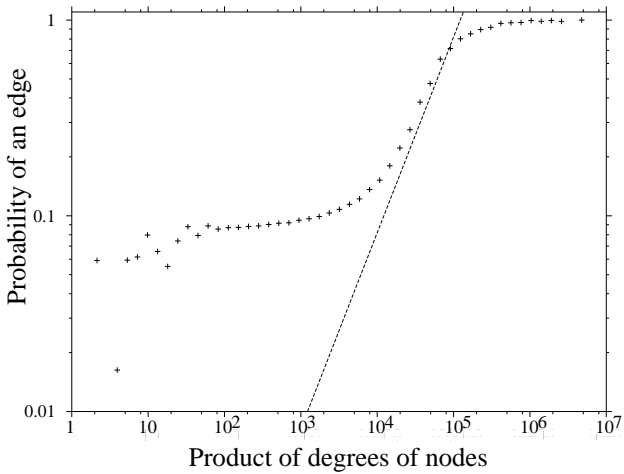}
\caption{Average probability of an edge between two nodes as a function of the product of the degrees of those nodes, calculated for LJ$_{14}$.
Values were calculated from an ensemble of random networks and are used to calculate $Q'$.
The number of edges predicted by this method is never larger than 1.
The straight line shows the value $k_ik_j/2M$ used in the original calculation of $Q$.
For the two highest degree nodes, this takes the value 33.46.}
\label{fig:prededges}
\end{figure}

The probability of an edge between two nodes is shown in Figure \ref{fig:prededges} plotted against the product of the degree of the two nodes for both the original prediction \(a_{ij}=k_ik_j/2M\) and $f_{ij}$ from rewiring, as used in calculating $Q'$.
$f_{ij}$ is roughly proportional to the product of the degrees of the nodes at either end of each edge for medium degrees, but there is a plateau at higher degrees in $f_{ij}$ so that $f_{ij} \le 1$ for all pairs of nodes.
$f_{ij}$ is then necessarily larger than $a_{ij}$ for nodes with lower degree since the sum over all pairs of nodes is $M$.
This approach has also proved useful for the assortativity coefficient \cite{Newman02,Doye04b}, which also requires the predicted number of edges between two nodes in a random network.

\begin{figure}
\includegraphics[width=8.4cm]{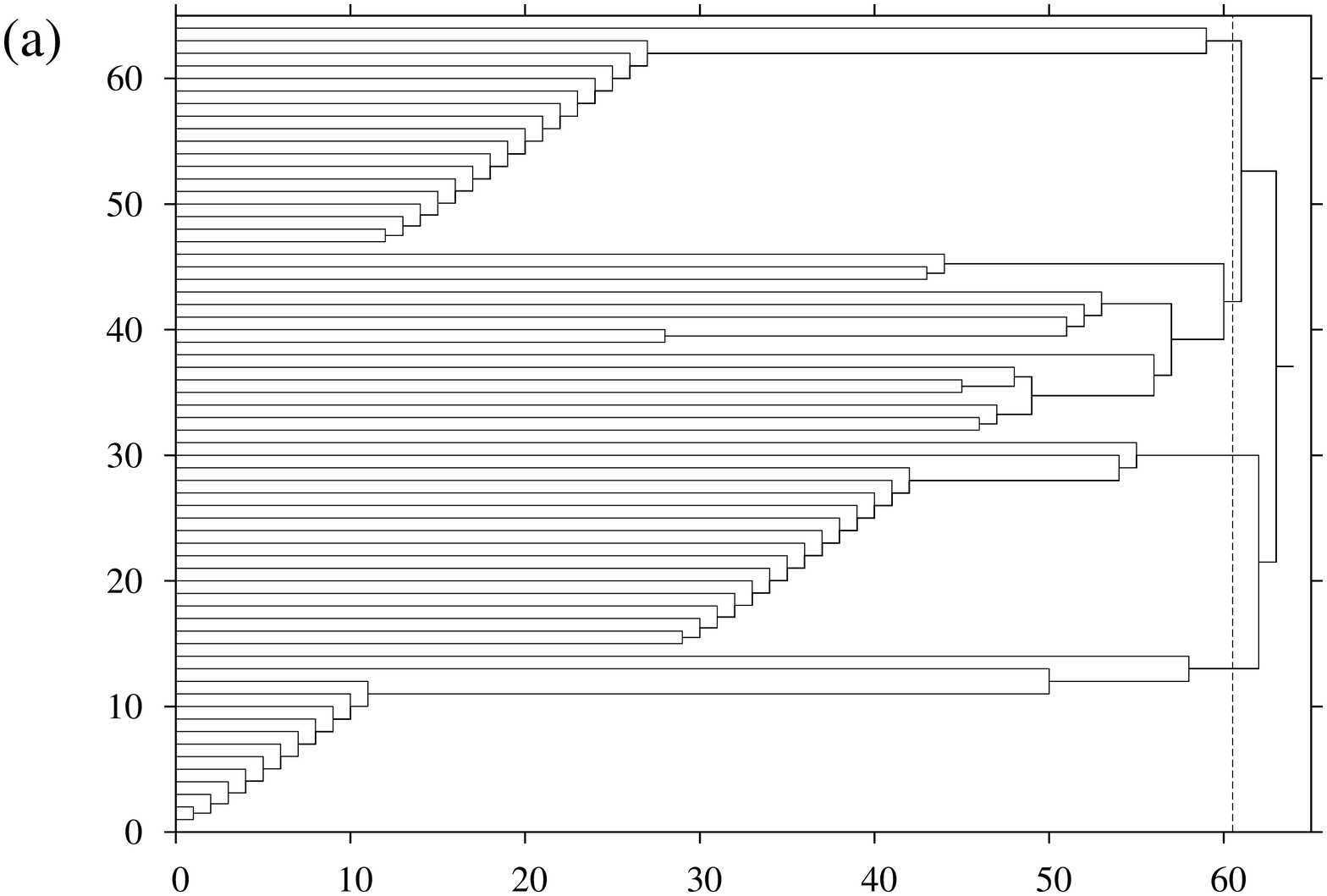}
\includegraphics[width=8.5cm]{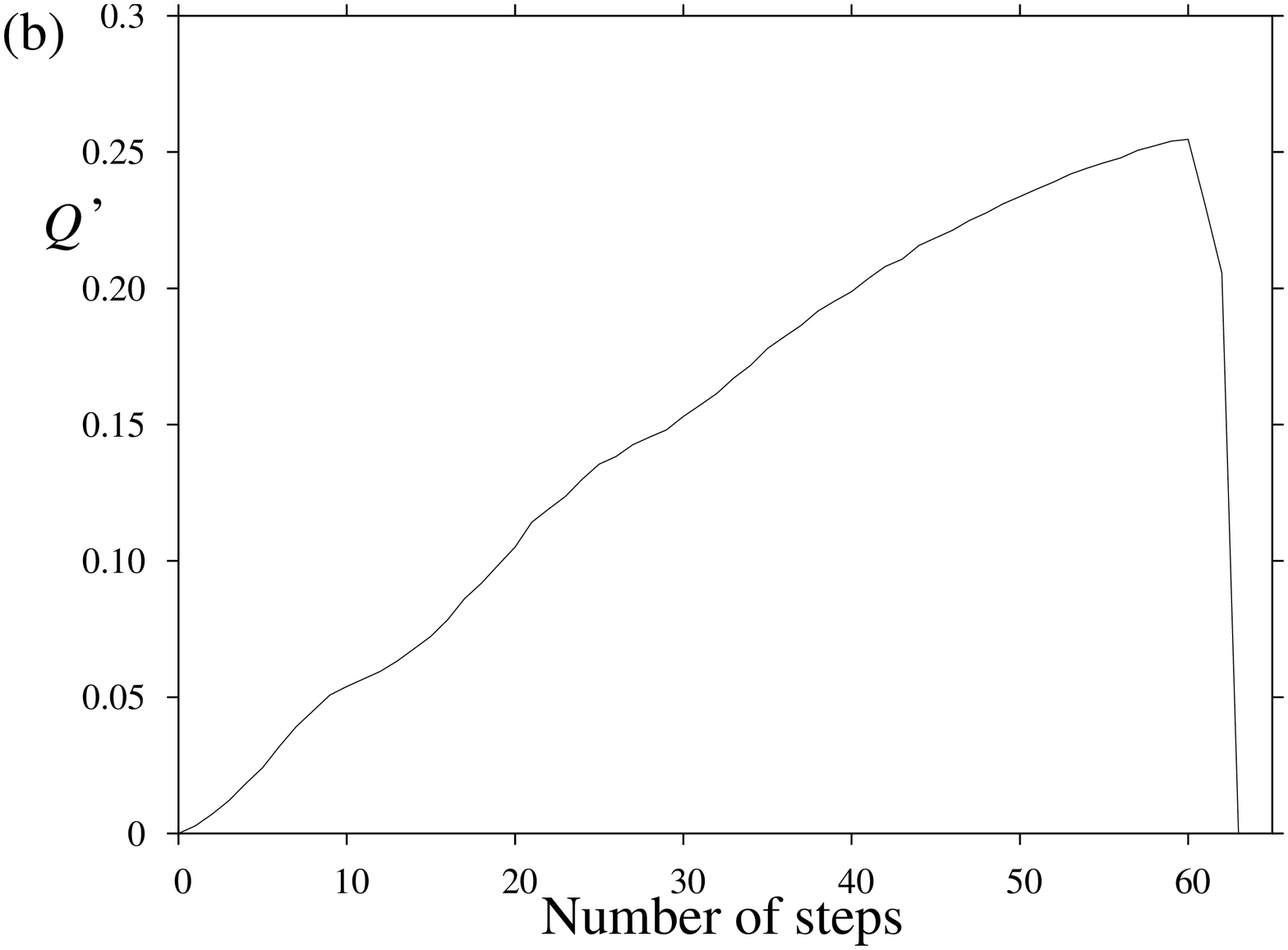}
\caption{(a) Dendrogram showing the results of the greedy algorithm using $Q'$ applied to LJ$_{10}$. The dashed line represents the best split found (with maximum $Q'$).
(b) Variation of $Q'$ as the algorithm progresses.}
\label{fig:fast-q'}
\end{figure}

Using $Q'$ rather than $Q$ will not affect the betweenness algorithm (although it could affect which community split is identified as the best), but it will almost certainly affect the greedy algorithm.
$Q'$ for the betweenness algorithm is shown in Figure \ref{fig:bet-bothq} and the dendrogram and $Q'$ for the greedy algorithm in Figure \ref{fig:fast-q'}.

The overall effect is unchanged; the greedy algorithm finds some community structure with a larger value of $Q'$ than for the betweenness algorithm, which finds essentially no community structure.
However, the community split found by the greedy algorithm is different for the two measures of modularity.
For the betweenness algorithm $Q'$ can be compared to $Q$ since both are measured for the same community splits.
The difference is significant for this network, with the ordering of the peaks being altered.
The highest peak of $Q'$ is further to the left because the original $Q$ disfavours small communities.

\begin{figure}
\includegraphics[width=8.6cm]{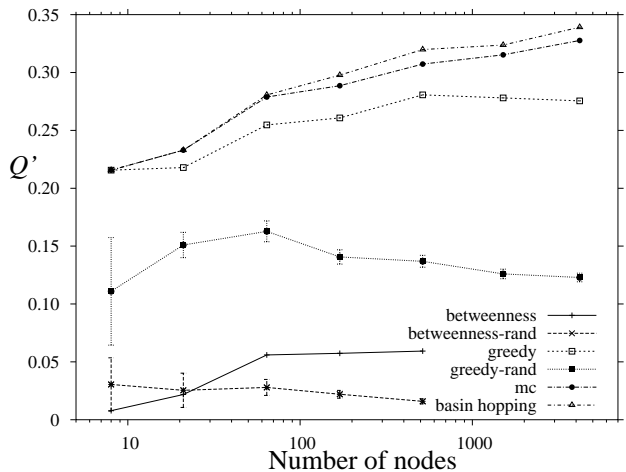}
\caption{$Q'$ against $N$ for cluster networks and random ensembles, error bars are one standard deviation from an ensemble of 25 random networks.
The results are (from bottom to top) for the betweenness algorithm, the greedy algorithm, simulated annealing and basin hopping. 
The betweenness algorithm was only run on clusters with up to 12 atoms (515 nodes) as it is slow for larger networks.}
\label{fig:QvsN}
\end{figure}

Higher values of $Q'$ were found by simulated annealing and basin hopping runs.
These values are shown in Figure \ref{fig:QvsN}, along with those for the greedy algorithm and the betweenness algorithm, for the real and random networks.
The modularity increases roughly logarithmically with the size of the networks.
The rate of increase for $Q'_{max}$ from the greedy algorithm slows down for larger networks, as seen previously \cite{Gronlund04}, simply because as the networks become larger the number of possible community splits increases and so the probability of finding the global maximum of $Q'$ decreases.
As the networks become larger, they also become more sparse ($p=2M/(N(N-1))$ decreases).
Therefore, there should be community splits with less edges between communities, and hence greater $Q'$.

$Q'$ is greater than zero for the random networks meaning that it is always possible to find some community structure.
This feature, coupled with the fact that $Q'$ depends on the size and average degree of the network \cite{Gronlund04}, implies that it is important to compare results with those for random networks \cite{Guimera04}.
$Q'_{max}$ is several standard deviations above the mean of the random networks, so although $Q'_{max}$ is fairly low compared to other networks, there is significantly stronger community structure than expected if nodes were connected randomly.
This is perhaps due to the spatial nature of the networks and their high clustering.
For example, previous work on the contact networks associated with models of space-filling disks (specifically the two-dimensional Apollonian packing) \cite{Doye04} found community structure with $Q_{max}=0.5938$.

Some correlations have been seen between community structure, high clustering and assortativity.
For example, these properties are mostly all seen in social networks, but rarely in technological or biological networks \cite{Newman03e}.
Gr\"onlund and Holme used a model from social network analysis to introduce community structure into an Erd\"os-Renyi random network for which clustering, assortativity and community structure (using the greedy algorithm) were then studied \cite{Gronlund04}.
The evolution of these properties was followed as the community structure was introduced.
Both $Q$ and the clustering coefficient increased from the Erd\"os-Renyi values and showed a strong correlation, fluctuating at the same times.
The networks became assortative, i.e.\,there were correlations between degrees of connected nodes (high degree nodes were more likely to be connected to other high degree nodes and low to low).
The assortativity coefficient fluctuated strongly, but differently to the clustering coefficient and the modularity, indicating that, although introducing community structure made the networks assortative, the link between the two properties is not as strong as that between community structure and clustering.

Newman studied a model scale-free network with community structure \cite{Newman03f} that was found to have a high clustering coefficient.
When different sized communities were present the model was also assortative by degree \cite{Newman03e}.
The explanation given for this was that nodes with high degree are more likely to be in big communities so that they can be in the same community as as many of their neighbours as possible.
Consequently, high degree nodes are likely to be in the same communities (the bigger ones) and are therefore also likely to be linked to each other.

\begin{figure}
\centerline{\includegraphics[width=8.6cm]{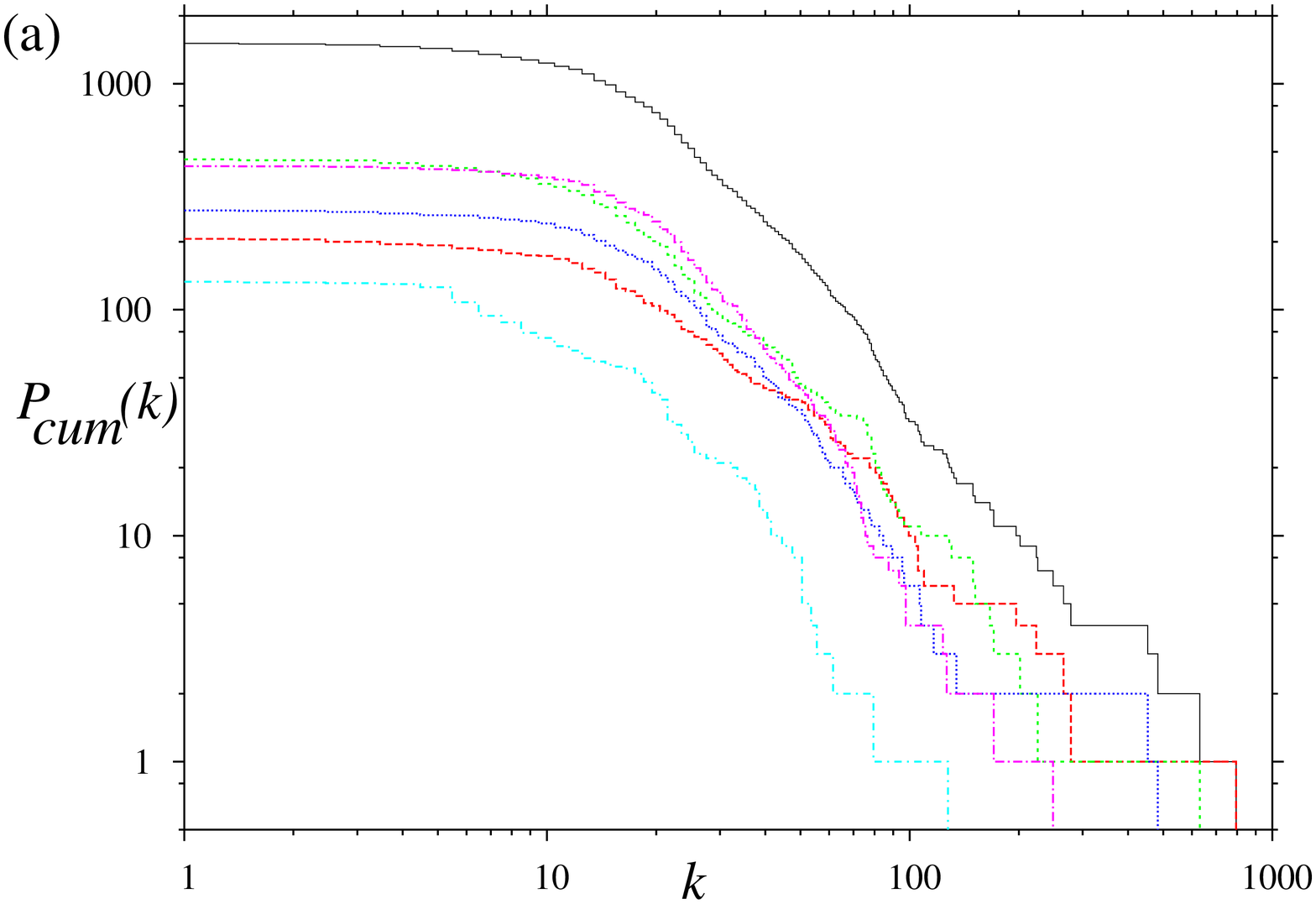}}
\centerline{\includegraphics[width=8.6cm]{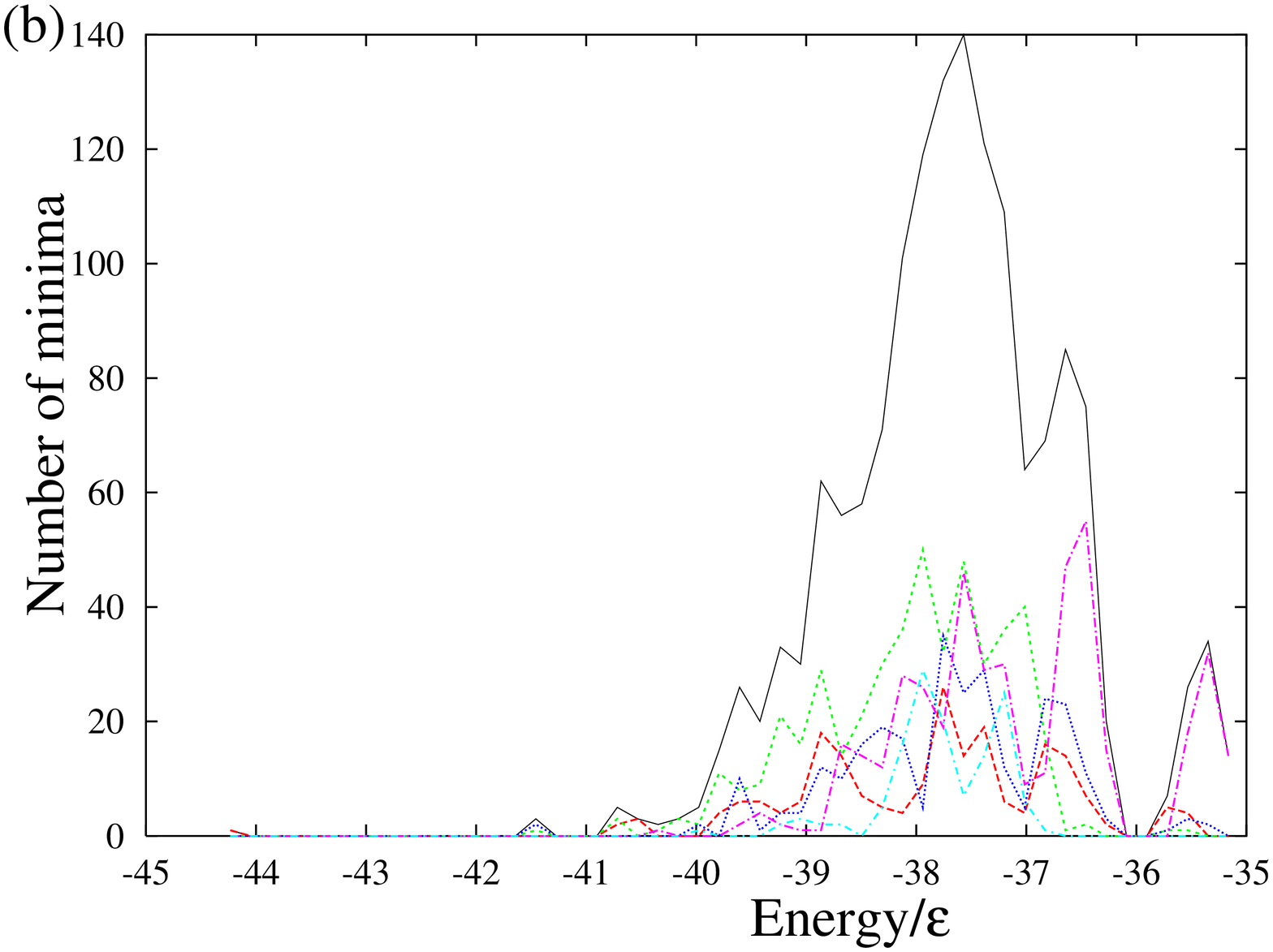}}
\caption{(Colour online) (a) Cumulative degree distribution and (b) energy distribution of minima for LJ$_{13}$. 
The solid (black) line is the overall distribution for the whole network and the dashed (coloured) lines those for the five communities found using basin hopping.}
\label{fig:pk-com}
\end{figure}

The networks studied here have been shown previously to be highly clustered \cite{Doye02}.
They are also disassortative (anticorrelated) by degree, but not significantly differently to random networks; this effect can be explained by the degree distribution and lack of multiple edges and self-connections \cite{Maslov04,Park03}.
The LJ networks have communities of differing sizes so they might also be expected to be assortative by degree if the hubs are mostly in big communities, whereas community sizes in the random networks vary less and as such would be expected to be less assortative.
As can be seen in plots of the degree distribution for each community (Figure \ref{fig:pk-com}) the hubs are split between three of the five communities for LJ$_{13}$.
These are not the three largest communities, the highest degree node in the second largest community (containing 432 nodes) has degree 248 whereas the biggest hub, with degree 794, is in a community of only 206 nodes.
This implies that there is some other property causing nodes to group into communities, beyond the degree and community sizes.

The networks are assortative with respect to their potential energy \cite{Doye04b}, i.e.\,minima with similar potential energies are likely to be connected.
It seems reasonable that nodes grouped in a community would correspond to minima with similar energies.
This is illustrated in the energy distribution shown in Figure \ref{fig:pk-com}.
There is some separation.
For example, most of the high energy minima are in one community, but the differences are fairly weak.
In a study of LJ$_{55}$ a similar overall energy histogram was seen and the different peaks could be assigned to different classes of structure \cite{Doye98}, namely different types and numbers of defect.
Previous work on 13-atom Morse clusters \cite{Miller99b} found that different peaks corresponded to structures with different numbers of non-nearest neighbours.
It is therefore likely that the different peaks in the energy histogram can also be differentiated by structure.

\subsection{BLJ$_{13}$}
A binary Lennard-Jones cluster with 13 atoms, one of which is distinguished as being heavy, was also studied.
The PEL has two funnels \cite{Landscapes,Wales04} leading to different isomers of the global minimum.
Networks for the clusters considered in the previous section were found in previous work \cite{Doye02b} using the eigenvector-following algorithm to locate all the stationary points.
The network for the binary system was obtained from this monoatomic 13-atom network in the following way.
For each minimum in turn, one atom was given a mass of two while the other twelve had a mass of one.
The trace of the inertia tensor was calculated as
\begin{equation}
\sum_{i} m_i \left( (x_i-x_{com})^2+(y_i-y_{com})^2+(z_i-z_{com})^2 \right)
\end{equation}
\noindent where the sum is over all the atoms, $m_i$ is the mass of atom $i$, $x_i$ its $x$ coordinate and $x_{com}$ the $x$-coordinate of the centre of mass.
This is repeated with a different atom being heavy until all 13 possibilities have been investigated.
The trace of the inertia tensor is used to distinguish different isomers, as the trace of the inertia tensor for some of the 13 possible isomers may be equal, depending on the symmetry of the cluster.
For example, the global minimum icosahedron with the heavy atom in any of the 12 outside positions are classed as indistinguishable.
There are a total of 17\,964 distinguishable minima (nodes) compared to 1509 for the monoatomic cluster.
For each distinguishable version of the transition states with one heavy atom, the associated pathway was found.
The original network tells us which minima a transition state connects, and the traces of the inertia tensors of the minima at either end of the pathway tells us which versions of those minima it connects.
This gave rise to 294\,285 edges in the network.
Due to the large size of the network, applying the rewiring method to create a random ensemble of networks is not feasible, so the original expression for $Q$ has been used.

\begin{figure}
\centerline{\includegraphics[width=8.6cm]{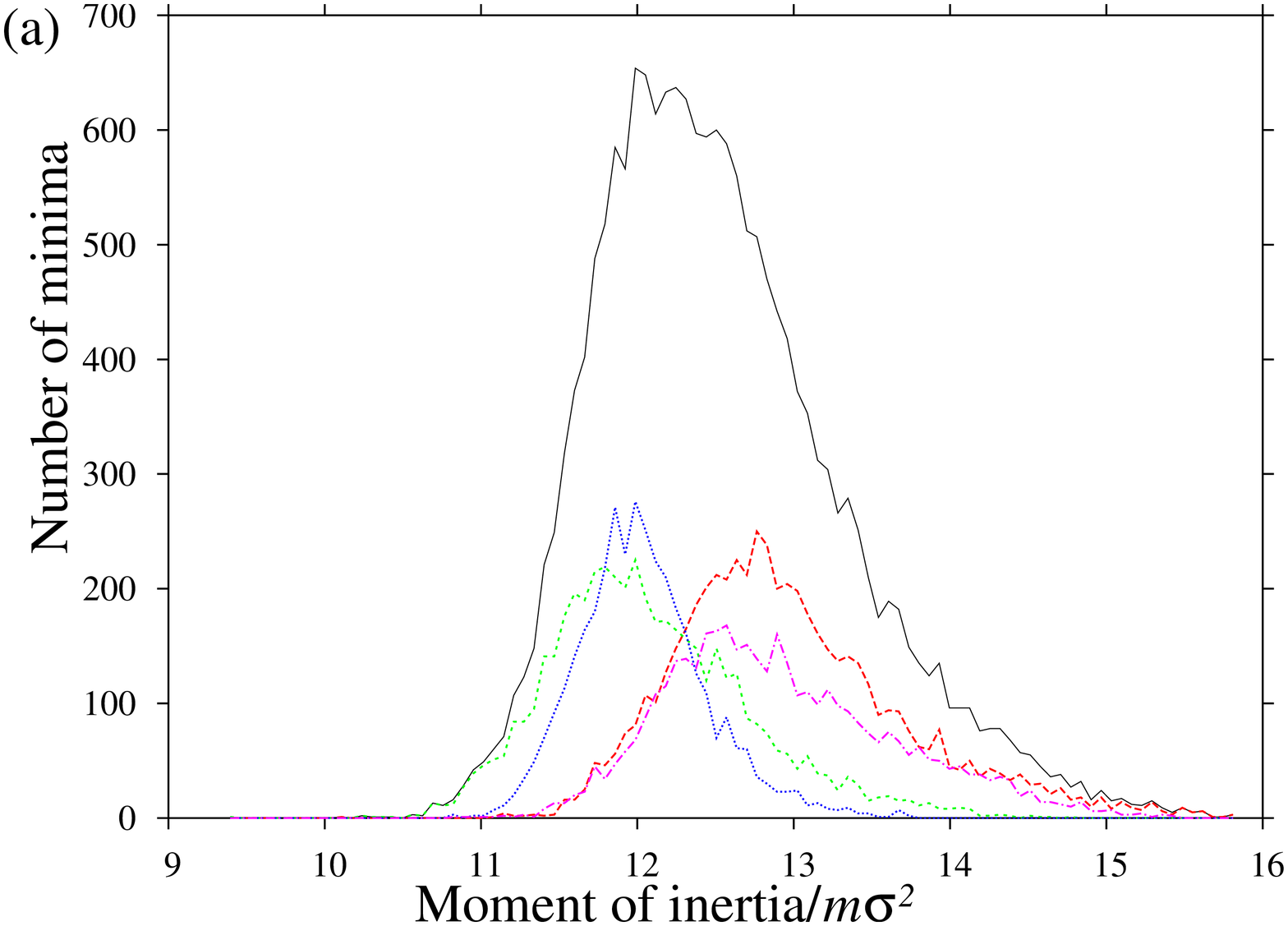}}
\vspace{5mm}
\centerline{\includegraphics[width=8.6cm]{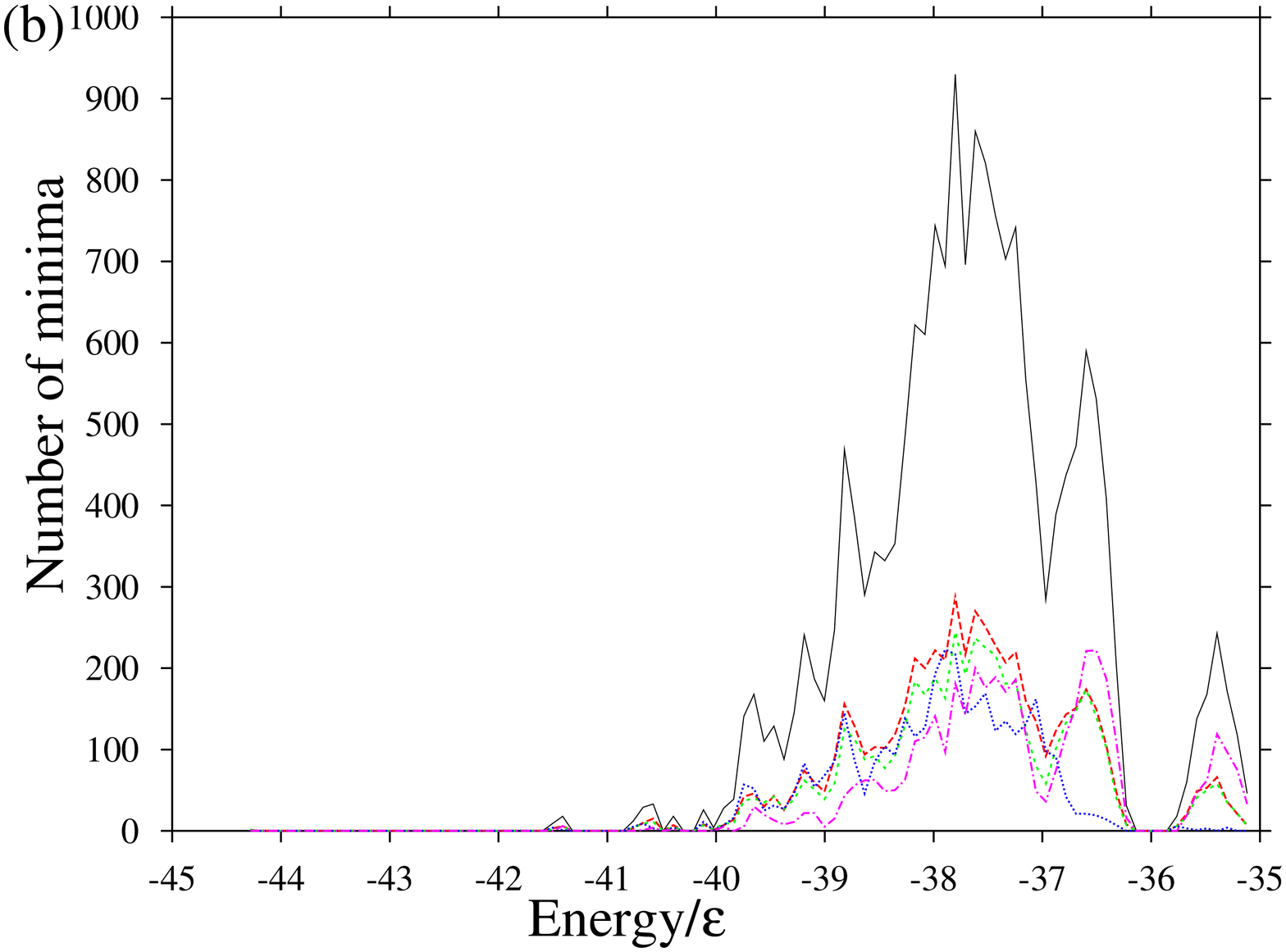}}
\caption{(Colour online) Distributions of (a) the moments of inertia (the geometric mean of the principal moments of inertia) and (b) energies of minima found for LJ$_{13}$ with one heavy atom, using basin hopping to optimize $Q$.
The solid (black) line shows the distribution for all minima.
The dashed (coloured) lines represent the four largest communities with sizes greater than 3000, the other communities all contained less than 20 nodes.
Minima with larger moments of inertia have the heavy atom closer to the outside of the cluster.}
\label{fig:hist-13fun}
\end{figure}

The 13-atom cluster with one heavy atom has higher modularity than the simple 13-atom cluster, the highest $Q_{max}$ value obtained was 0.4370 using the basin hopping algorithm.
This reflects the greater heterogeneity of this PEL compared to those of the smaller clusters.
The two permutational isomers of the global minimum are in different communities, the two largest (consisting of 5507 and 4732 nodes).
Measuring the moment of inertia (the geometric mean of the 3 principal moments of inertia) for the clusters gives an indication of where the heavy atom is; i.e.\,if it is near the edge of the cluster the moment of inertia will be larger and \textit{vice versa}.
The distributions of the moments of inertia are shown in Figure \ref{fig:hist-13fun} for the four largest communities.
Minima in the largest community, containing the version of the global minimum with the heavy atom on the outside, have high inertia implying that the other minima also have the heavy atom close to the outside.
The converse is true for the second largest community, which contains the permutational isomer of the global minimum with the heavy atom in the centre.
The energy distributions (Figure \ref{fig:hist-13fun}) shows that these two communities have very similar energy distributions and therefore probably consist of different permutational isomers of the same geometric structures.
There are two additional communities which are fairly large (4008 and 3629 nodes).
These also have peaks in their inertia distributions at around the same position as the two largest communities, implying that the two funnels around the two permutational isomers of the global minimum have been broken down into smaller communities.
One of the two smaller communities consists of fairly high energy minima and could be a transition region between the two funnels.

\subsection{LJ$_{38}$}
\begin{figure}
\centerline{\includegraphics[width=8.6cm]{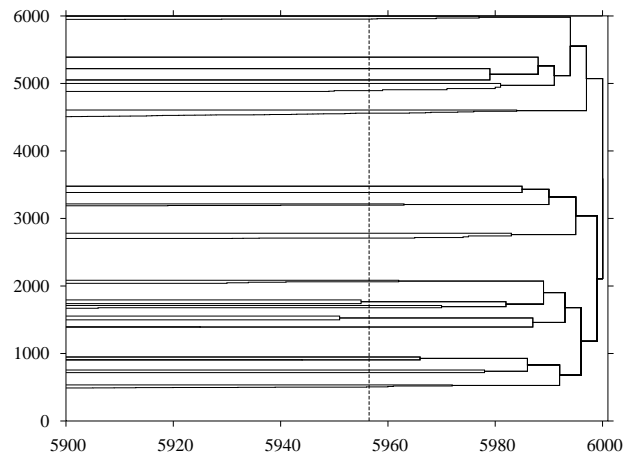}}
\caption{The final 100 steps in the dendrogram for LJ$_{38}$ using the greedy algorithm.
If a community only consists of single nodes, i.e.\,is not composed of smaller communities, then for clarity only the first and last nodes to join that community are shown.
The variation of $Q'$ as the algorithm progresses is not shown, it has a single peak of value $Q'_{max}$=0.8311 at level 5956 (indicated by a dashed line on the dendrogram).}
\label{fig:38dend}
\end{figure}

For LJ$_{38}$ the community structure is very strong with $Q'_{max}$ from the greedy algorithm being 0.8311, which is of the order of four times that found for the smaller LJ networks and one of the higher modularities found for any network.
$Q'$ tends to be higher than $Q$ as the lack of self-connections means too many edges are generally predicted within communities for $Q$, but $Q_{max}$ is also greater than 0.8.
This indicates that the PEL for LJ$_{38}$ is much more heterogeneous than those for the smaller clusters, as expected.
$Q'_{max}$ was also high for the random ensemble with a mean of 0.5650 implying that the degree distribution can explain some of the community structure.
The degree distribution of the minima sampled for this cluster is not scale-free, because of the incompleteness of the network.
In a scale-free network there are some nodes with very high degree (hubs).
It is not always likely for a hub to be in the same community as all of its neighbours, for example in LJ$_{14}$ the largest hub is connected to approximately 75\,\% of the nodes in the network and there is no community that large.
This could mean that community structure can be stronger in networks with degree distributions without hubs, such as that for LJ$_{38}$, explaining the high values of the modularity seen for the random ensemble.
However, the value of $Q'_{max}$ for the PEL network is over 17 standard deviations higher than that for the random networks and so is significant.
The dendrogram is shown in Figure \ref{fig:38dend}, only the last 100 of 6000 steps are shown.

The community structure found can be compared to the multiple-funnel structure of the PEL.
Some of the minima have previously been assigned to either an fcc or icosahedral funnel using a master equation approach.
At the point of maximum $Q'$ in the community structure found from the greedy algorithm, the minima from the fcc funnel are in the same community and there are no icosahedral structures in that community.
This is also true for the partition with the highest value of $Q'$ ($Q'$=0.8314), obtained from basin hopping.
The best split is at level 5956 ($Q'_{max}$=0.8311) and the fcc and icosahedral minima are in different communities until level 5990 where $Q'$=0.8099 and has begun to fall off.
The betweenness algorithm is too slow to run to completion, but after removal of only 13 edges the network was split into two communities with all fcc minima in one and all icosahedral minima in the other.
The community algorithms are therefore giving an insight into the topology of the PEL whilst only using data on the connectivities of the minima, whereas the previous methods also use the energies of minima and the barrier heights between them.
This should be contrasted with the community structure found for the smaller clusters, which is very weak and was not found at all using the betweenness algorithm.
The PELs for the small clusters have a single funnel topography. 
The 13-atom Morse cluster, which is similar to LJ$_{13}$ when appropriate parameters are used, has only one monotonic sequence basin, meaning that a path from any minimum which decreases the energy at each step leads to the global minimum \cite{Doye99b}.
This information from topographical analysis is consistent with the picture obtained from the topological approach in this work.

\subsection{2D hexagonal lattice}

Since the energy landscapes of small clusters have single funnel topographies, it is curious that there should be any community structure in the corresponding networks.
It is possible that this community structure is related to the spatial nature of the landscapes.
Nodes in the same region of space are more likely to be interconnected (leading to high clustering).
If these nodes are grouped together into communities, there is likely to be many edges within those communities and few between different communities, as edges connecting different communities would only come from nodes adjacent to the boundaries.
To investigate the effect of spatial organization and clustering on community structure a hexagonal lattice was studied.

The greedy algorithm can be applied to a large two-dimensional hexagonal lattice, where all nodes have the same degree ($k$=6) and the clustering coefficient is 0.4.
In the first step any edge is equally likely to be added, each changing $Q$ to $Q + \Delta Q$, where $\Delta Q = \frac{(1-k^2/2M)}{M}$.
If $k^2/2M < 1$, i.e.\,$M > 18$ then $\Delta Q$ is positive.
In the second step one of the two nodes that forms a triangle with that edge will be added, with $\Delta Q = 2 \times \frac{(1-k^2/2M)}{M}$ (also positive if $M > 18$).
In the third step, a fourth node will be added to the community.
This can only increase the number of edges within the community by two, whereas the increase in the number of expected edges will be the number of new pairs of nodes (the size of the community, $n_c = 3$) multiplied by the probability of an edge between each pair, $\Delta Q = \frac{ 2 - n_c k^2/2M}{M}$.
In the following steps, the number of edges added to the community can be increased by up to 6 (the degree of each node) depending on the shape of the growing community (this will depend on the random choices).
At each step the expected number of edges (to be subtracted) is $n_c k^2/2M$.
At some point, $\Delta Q$ will be smaller than that for starting a new community ($\Delta Q = \frac{(1-k^2/2M)}{M}$) so a new community will be started.

\begin{figure}
\centerline{\includegraphics[width=8.6cm]{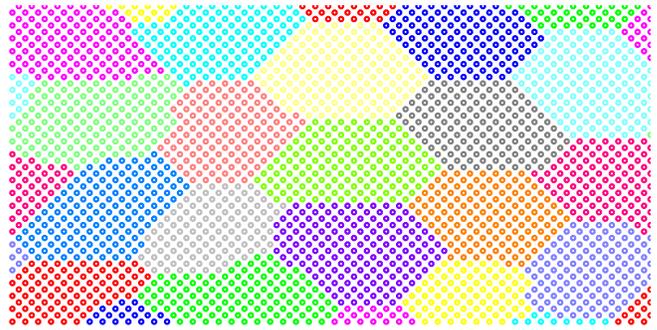}}
\caption{(Colour online) The partition with the highest modularity ($Q'$=0.8480) seen for a two-dimensional hexagonal lattice with 2500 nodes using periodic boundary conditions.
This partition was found using basin hopping to optimize $Q'$ from an initial partition into roughly equally sized hexagons.}
\label{fig:lat}
\end{figure}

$Q$ and $Q'$ were optimized for a hexagonal lattice with periodic boundary conditions such that all nodes have degree 6.
The following results correspond to the largest system studied, a lattice with 2500 nodes.
These were also compared to a random ensemble.
$Q_{max}$ and $Q'_{max}$ from greedy optimization are 0.7674 and 0.7231 respectively, whereas the corresponding values for the random networks are 0.3860 and 0.3517, differences of 191 and 45 standard deviations.
The highest value of $Q'$ (0.8480) corresponds to the partition shown in Figure \ref{fig:lat}, which was found by basin hopping from a partition into roughly equally sized hexagons ($Q'$=0.8389 for 18 communities of average size 139).
Interestingly, the partition obtained from greedy optimization contained much larger communities, with an average size over 300.
Nodes in the lattice have constant degree, so this could possibly explain why the modularity is higher than for scale-free landscape networks (as for LJ$_{38}$), for both the lattice and the randomized versions.
However, Apollonian networks, which are spatial scale-free networks, also have a high modularity.
The modularity of a scale-free network embedded on a lattice could be studied \cite{Rozenfeld02} to provide a further comparison.

The modularity of the lattice is still much higher than that for the random ensemble so the community structure is significant.
Nodes close together on a hexagonal lattice have many edges between them due to the highly clustered nature of the network.
They also have few edges to the rest of the network, as they only occur at the edges of the community.
Similar high values of the modularity have been analytically predicted for low-dimensional regular lattices \cite{Guimera04}.
The partition of the lattice into communities is highly degenerate, for example the partitions could be translated in any direction, giving a different community split with the same high value of the modularity.
This raises questions about the interpretation of $Q$ and $Q_{rand}$ in terms of `unique' communities.
Other methods for finding community structure have been proposed which give many possible partitions due to random choices in the algorithms \cite{Reichardt04,Wilkinson04}.
These partitions could then be compared to determine whether they are based on a strong community structure or are very different to each other, as would be the case for the hexagonal lattice.

An alternative way to probe this issue would be to look at the thermodynamics associated with partitioning of the network, where $-Q$ or $-Q'$ plays the role of energy, as in the Monte Carlo methods used here.
If there was a unique, well-defined community structure, one would expect a transition, somewhat akin to crystallization, to a low-temperature state with a low density of partitions, all of which would have the same basic structure.
By contrast, a more `glass-like' behaviour would be expected for networks with ill-defined community structure.
That is there would only be a gradual increase in $Q$ as the temperature is decreased without any sharp transitions, and there would be many high-modularity partitions with significant structural differences.

\section{Conclusion}

The potential energy landscapes of small Lennard-Jones clusters have been studied in terms of networks describing which metastable states of the clusters are connected.
A number of algorithms have been used to uncover the community structure of the PEL networks.
All are based solely on the topology of the network, i.e.\,barrier heights and energies of the minima are not taken into account.
The first algorithm used, based on edge betweenness, found no community structure in the networks of the small clusters, consistent with a single funnel shape of the PEL.
However, this algorithm has been unsuccessful in previous studies when the network is densely connected, and the PEL networks studied here have a higher average degree than many of the other networks previously studied.
The other algorithms put nodes into communities such that there are more edges within communities than expected for a random network, thus optimizing the modularity.
A modification was made to the way in which the community structure is quantified.
$Q'$ is essentially the same as $Q$, measuring the fraction of edges within communities compared to the predicted fraction for a random network with the same degree distribution.
However, here the random network does not contain multiple edges or self-connections making it a better comparison.

Optimizing the modularity did find weak community structure in the networks.
This implies that within the single funnel of the PEL minima form groups with more transition states between them than to minima in other groups.
The community structure is likely to be connected to the spatial nature of the networks and therefore related to the clustering.
To investigate the effect of spatial ordering of a network on its community structure a hexagonal lattice was studied.
Strong community structure was found for the lattice, implying that the weak community structure found in the landscape networks could be due to spatial organization.

Two LJ clusters with more complicated PELs where community structure was expected were also studied.
Labelling one atom in LJ$_{13}$ distinguishes permutational isomers that fit broadly into two classes, those with the heavy atom on the outside and those with the heavy atom on the inside.
Two communities corresponding to these classes of structure were found by optimizing $Q$.
The PEL of LJ$_{38}$ is known to have two funnels, one consisting of fcc structures containing the truncated octahedron global minimum and the other consisting of icosahedral structures.
Both topological algorithms found much stronger community structure for this cluster than for the smaller clusters, implying a much more heterogeneous PEL.
The community structure found was also compared to the known funnel structure of the PEL.
In both algorithms fcc and icosahedral structures were in different communities.
The topology of the PEL therefore contains information about its heterogeneity and can be used to provide a similar picture to that gained from a topographical analysis.


\end{document}